\documentclass[%
 reprint,
 amsmath,amssymb,
 aps,
]{revtex4-2}

\usepackage{xcolor}
\usepackage{graphicx}% Include figure files
\graphicspath{{./Figs/}}
\usepackage{dcolumn}% Align table columns on decimal point
\usepackage{bm}% bold math
\begin{document}
\preprint{APS/123-QED}

\title{Retraction of levitating drops}% Force line breaks with \\

\author{Kindness Isukwem}
\thanks{kindness-chinwendu.isukwem@minesparis.psl.eu}
\affiliation{Mines Paris, PSL University, Centre for material forming (CEMEF), UMR CNRS 7635, rue  Claude Daunesse, 06904 Sophia-Antipolis, France}

\author{Elie Hachem}
\thanks{elie.hachem@minesparis.psl.eu}
\affiliation{Mines Paris, PSL University, Centre for material forming (CEMEF), UMR CNRS 7635, rue  Claude Daunesse, 06904 Sophia-Antipolis, France}%

\author{Anselmo Pereira}
\thanks{anselmo.soeiro\_pereira@minesparis.psl.eu}
\affiliation{Mines Paris, PSL University, Centre for material forming (CEMEF), UMR CNRS 7635, rue  Claude Daunesse, 06904 Sophia-Antipolis, France}%

\date{\today}% It is always \today, today,

%%%%%%%%%%%%%%%%%%%%%%%%%%%%%%%%%%%%%%%%%%%%%%%%%%%%%%%%%%%%%%%%%%%%%%%%%%%%%%
%%%%%%%%%%%%%%%%%%%%%%%%%%%%%%%%%%%%%%%%%%%%%%%%%%%%%%%%%%%%%%%%%%%%%%%%%%%%%%
\begin{abstract}
This theoretical and numerical study focuses on the physical mechanisms driving the retraction of levitating Newtonian micrometric/millimetric/centimetric drops surrounded by air and under zero-gravity conditions. The drops present a pancake-like initial shape, gradually converging towards a spherical one under surface tension effects. Three drop retraction regimes are observed: capillary-inertial; mixed capillary-inertio-viscous; and capillary-viscous. In the first regime, the retraction is essentially driven by a competition between capillary pressure and inertial stresses, which induces a complex flow with equivalent shear, uniaxial and biaxial components. As the viscous stress becomes comparable to the capillary and the inertial stresses, the second regime emerges while shear-based deformations tend to vanish. Lastly, the third regime is dominated by a balance between capillary and viscous stresses, essentially exhibiting axial deformation. These physical features are underlined through multiphase three-dimensional numerical simulations and analysed in light of retraction dynamics, energy transfer and scaling laws. Our results are rationalised in a two-dimensional diagram linking the drop retraction time with the observed retraction regimes through a single dimensionless parameter combining capillary, inertial, viscous and geometrical effects, i.e., the \textit{retraction number}.   
\end{abstract}
%%%%%%%%%%%%%%%%%%%%%%%%%%%%%%%%%%%%%%%%%%%%%%%%%%%%%%%%%%%%%%%%%%%%%%%%%%%%%%
%%%%%%%%%%%%%%%%%%%%%%%%%%%%%%%%%%%%%%%%%%%%%%%%%%%%%%%%%%%%%%%%%%%%%%%%%%%%%%

\maketitle

%%%%%%%%%%%%%%%%%%%%%%%%%%%%%%%%%%%%%%%%%%%%%%%%%%%%%%%%%%%%%%%%%%%%%%%%%%%%%%
%%%%%%%%%%%%%%%%%%%%%%%%%%%%%%%%%%%%%%%%%%%%%%%%%%%%%%%%%%%%%%%%%%%%%%%%%%%%%%
%--------------------------------------------------------------------------------------------------------------------------------------------------------------------------------------%--------------------------------------------------------------------------------------------------------------------------------------------------------------------------------------
\section{Introduction} \label{INTRO}
%--------------------------------------------------------------------------------------------------------------------------------------------------------------------------------------%--------------------------------------------------------------------------------------------------------------------------------------------------------------------------------------
The deformation of dispersed filaments, sheets and drops in a suspending continuous fluid is a paramount problem in Fluid Mechanics \citep{Dupre-67, Dupre-68, Rayleigh-91, Taylor-32, Taylor-34, Taylor-64, Grace-82, Rallison-84, Stone-94, Eggers-08b, Minale-10, Eggers-20, Pierson-20, Deka-20, Lohse-22, Sanjay-23, Ni-23}. Several studies on this topic have focused on the spreading and fragmentation of drops impacting solid surfaces \citep{Villermaux-07, Villermaux-11, Josserand-16}, liquids \citep{Jalaal-19, Tang-19}, and gas interfaces \citep{Adda-Bedia-16, Isukwem-24d}, as well as on the stretching and breakup of drops within shear flows and contractions \citep{Stone-86, Tjahjadi-91, Stone-94, Vervoort-05, Tregouet-18, Crialesi-Esposito-23, Ni-23}. These studies hold a cardinal position in numerous applications, including the architecting of complex fluids \citep{Leal-Calderon-07, Visser-18}, mixing optimization \citep{Briscoe-99, Garstecki-06, Salonen-20}, spray generation \citep{Soto-18, Villermaux-22}, atomization \citep{Gorokhovski-08}, firefighting \citep{Lorenceau-03, Ryu-17, Kumar-18}, inkjet printing \citep{Lohse-22}, encapsulations of bioactive agents \citep{Alessandri-13, Keymeulen-23}, microfluidics \citep{Anna-16, Dewandre-20}, drag reduction \citep{Ceccio-10, Ezeta-19} and 3D printing of electronic components, organs, tissues, and prosthetics \citep{Murphy-14, Modak-20}.   

A cardinal examination concerns the relaxation of prolate drops embedded in a viscous liquid matrix after strain jumps \citep{Acrivos-78, Bentley-86, Hinch-80, Stone-89, Stone-94, Guido-99}. This inertia-free process is triggered by surface tension effects and is highly affected by the morphology of the drop and the viscosity of the involved fluids \citep[dispersed and continuous phases; ][]{Assighaou-08, Renardy-09, Becu-09, Assighaou-10}. 

\begin{figure*}%[h!]
\centering
\includegraphics[angle=0, scale=0.281]{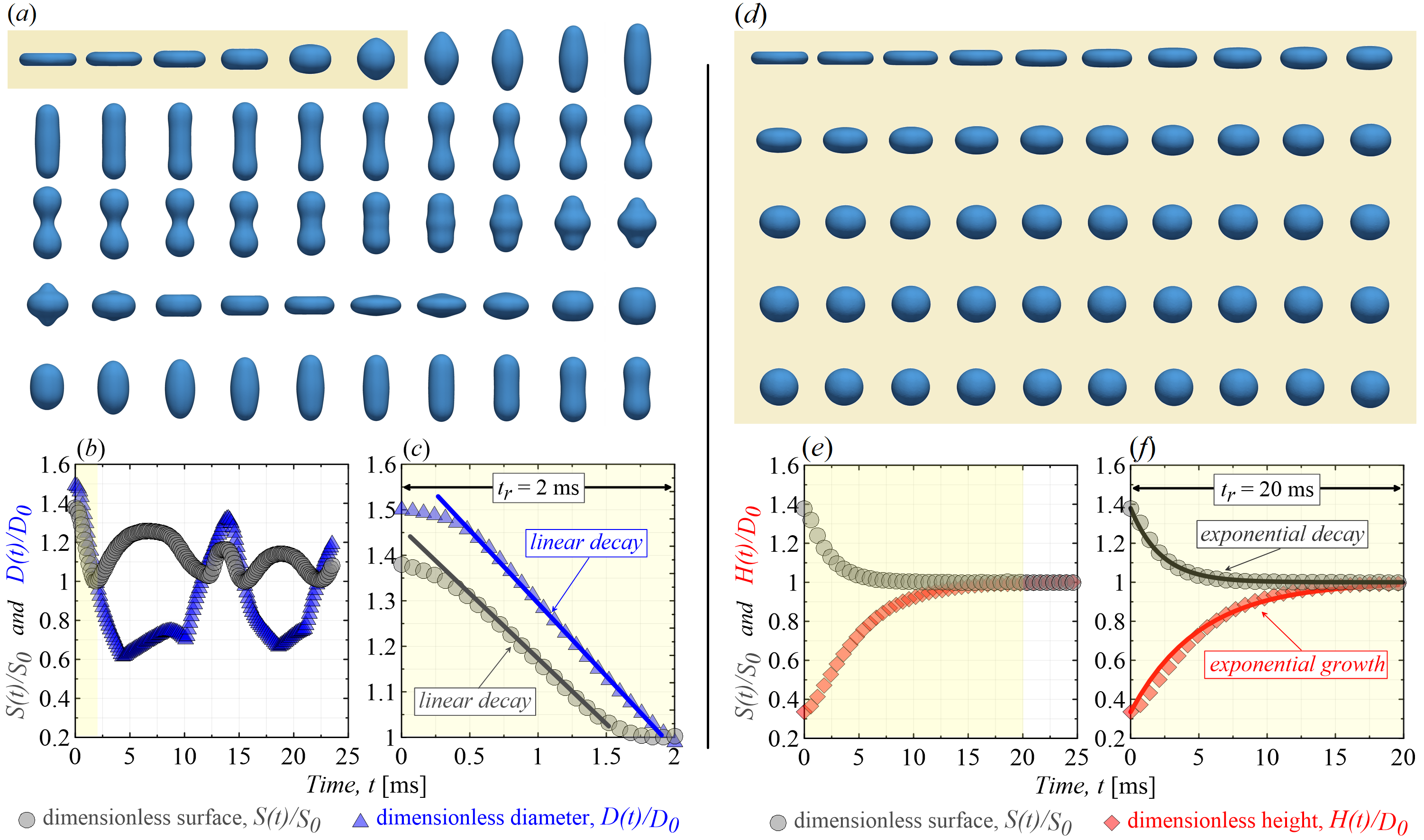}
\vspace*{-0.2cm}
\caption{Retraction dynamics in levitating drops (zero-gravity conditions). (\textit{a}) Numerical snapshots of a pancake-like drop of initial diameter $D_{max} =$ 9mm, initial thickness $H_{min} =$ 2mm, $\rho =$ 100kg/m$^3$, $\sigma =$ 0.2N/m and $\eta =$ 0.000905Pa$\cdot$s oscillating multiple times between the pancake-like, spherical and the filament-like shape before achieving its zero-velocity spherical form (not shown for brevity) whose radius is $R_0$ and surface is $S_0 = 4\pi R_0^2$. (\textit{b}) Its instantaneous dimensionless surface $S(t)/S_0$ (grey circles) and diameter $D(t)/D_0$ (blue triangles) are also plotted against time $t$. (\textit{c}) Zoomed plot of the retraction interval $t_r$ (i.e. yellow regions in \textit{a} and \textit{b}), highlighting the linear decay of $S(t)/S_0$ and $D(t)/D_0$ with $t$. (\textit{d}) Numerical snapshots of a levitating drop of initial diameter $D_{max} =$ 9mm, initial thickness $H_{min} =$ 2mm, $\rho =$ 100kg/m$^3$, $\sigma =$ 0.2N/m and $\eta =$ 0.12Pa$\cdot$s during retraction without oscillations. (\textit{e}) Its instantaneous dimensionless surface $S(t)/S_0$ (grey circles) and height $H(t)/D_0$ (red squares) are plotted against time $t$. (\textit{f}) Zoomed plot of the retraction interval $t_r$ (i.e. yellow regions in \textit{d} and \textit{e}), underlining the monotonic exponential decay of $S(t)/S_0$ and growth of $H(t)/D_0$ with $t$. The interval $\Delta t$ between two snapshots (in \textit{a} and \textit{d}) is 0.4ms.}
\label{fig-1}
\end{figure*}

Another key but scarcely conducted examination concerns capillary-induced retraction of stretched drops of viscosity $\eta$, density $\rho$ and surface tension $\sigma$, surrounded by air and under negligible gravitational effects conditions. The retraction of levitating objects can be observed when analysing Leidenfrost drops \citep{Leidenfrost-56, Hall-69, Garmett-78, Maquet-16, Adda-Bedia-16, Gauthier-19, Graeber-21, Wang-22, Isukwem-24d}, bouncing drops \citep{Jayaratne-64, Gopinath-01, Richard-02, Bach-04, Zou-11, Molacek-13a, Tang-19}, walking drops \citep{Couder-05, Molacek-13b, Bush-20}, and levitating drops in acoustic fields \citep{Hasegawa-19, Chen-24}, for instance. As illustrated by the numerical snapshots displayed in figures \ref{fig-1}(\textit{a}) and \ref{fig-1}(\textit{d}), depending on the balance between capillarity, inertia and viscosity, a zero-velocity stretched pancake-like drop can retract following at least two distinct dynamics. In the first scenario, shown in sub-figures \ref{fig-1}(\textit{a})-(\textit{c}), the pancake-like object of initial diameter $D_{max} =$ 9mm, initial thickness $H_{min} =$ 2mm, $\rho =$ 100kg/m$^3$, $\sigma =$ 0.2N/m and $\eta =$ 0.000905Pa$\cdot$s oscillates multiple times between the pancake-like, spherical and the filament-like shape before achieving its zero-velocity spherical form (not shown for brevity) whose radius is $R_0$ and surface is $S_0 = 4\pi R_0^2$. The retraction dynamics is also highlighted by the time-evolution of the surface $S(t)$ and the diameter $D(t)$ of the drop (made dimensionless by $S_0$ and $D_0 = 2R_0$, respectively) shown in sub-figures \ref{fig-1}(\textit{b})-(\textit{c}). Interestingly, from the very beginning of the process until the drop first reaches $S_0$, in the interval called here \textit{retraction time} and denoted by $t_r$ (underlined by the yellow regions), a linear decay is observed, followed by an oscillatory movement. The linear decay is better perceived in sub-figure \ref{fig-1}(\textit{c}) in which $0 \leq t \leq t_r$. In the second retraction dynamics, the stretched drop, which is about 100 times more viscous ($\eta =$ 0.12Pa$\cdot$s) than that illustrated in sub-figures \ref{fig-1}(\textit{a})-(\textit{c}), slowly and exponentially evolves towards the spherical shape while viscous effects dissipate entirely its kinetic energy. This process is highlighted by the snapshots in sub-figure \ref{fig-1}(\textit{d}), and the dimensionless surface $S(t)/S_0$ and height $H(t)/D_0$ plotted in sub-figures \ref{fig-1}(\textit{e})-(\textit{f}) as a function of time $t$ (these profiles will be discussed deeply in section \ref{RD}). Moreover, it is also essential to observe that in sub-figure \ref{fig-1}(\textit{d})-(\textit{f}) $t_r$ appears ten times higher than in sub-figures \ref{fig-1}(\textit{a})-(\textit{c}) due to the increase of the drop viscosity.     

The shift from linear to exponential profiles brought out by figure \ref{fig-1}, as well as the increase of $t_r$, suggests the existence of different retraction regimes. In these connections, the following open question cannot be overemphasized: can one highlight the physical mechanisms driving the retraction dynamics of levitating drops, underlining the link between the properties and the morphology of the stretched objects with $t_r$? This question will be addressed in the present study.   

Different from the majority of the recently reported works \citep[see for instance][]{Villermaux-07, Savva-09, Conto-19, Pierson-20, Deka-20, Wang-23, Sanjay-23}, which considers sheets and filaments, we present here a theoretical and numerical investigation devoted to the physical mechanisms driving the capillary-induced retraction of levitating Newtonian micrometric/millimetric/centimetric drops surrounded by air in zero-gravity conditions. We only consider the drop retraction taking place in the interval between $t/t_r = 0$ and $t/t_r = 1$ ($0 \leq t \leq t_r$; highlighted by the yellow regions in figure \ref{fig-1}). Physical mechanisms are stressed using multiphase three-dimensional (3D) numerical simulations and analysed in light of retraction dynamics, energy transfer and scaling laws. Our results are rationalised in a two-dimensional diagram linking the drop retraction time with the observed retraction regimes through a single dimensionless parameter combining capillary, inertial, viscous and geometric effects, i.e., the \textit{retraction number}. 

The organization of the paper is as follows. A detailed description of the physical formulation and the used numerical method is presented in section \ref{PFNMDN}. Numerical results and theoretical arguments are discussed in section \ref{RD}. Finally, conclusions and perspectives are drawn in the closing section.

%%%%%%%%%%%%%%%%%%%%%%%%%%%%%%%%%%%%%%%%%%%%%%%%%%%%%%%%%%%%%%%%%%%%%%%%%%%%%%
%%%%%%%%%%%%%%%%%%%%%%%%%%%%%%%%%%%%%%%%%%%%%%%%%%%%%%%%%%%%%%%%%%%%%%%%%%%%%%
\section{Physical Formulation, Numerical Method and Dimensionless Numbers} \label{PFNMDN}
%%%%%%%%%%%%%%%%%%%%%%%%%%%%%%%%%%%%%%%%%%%%%%%%%%%%%%%%%%%%%%%%%%%%%%%%%%%%%%
%%%%%%%%%%%%%%%%%%%%%%%%%%%%%%%%%%%%%%%%%%%%%%%%%%%%%%%%%%%%%%%%%%%%%%%%%%%%%%

\begin{figure*}%[h!]
\centering
\includegraphics[angle=0, scale=0.275]{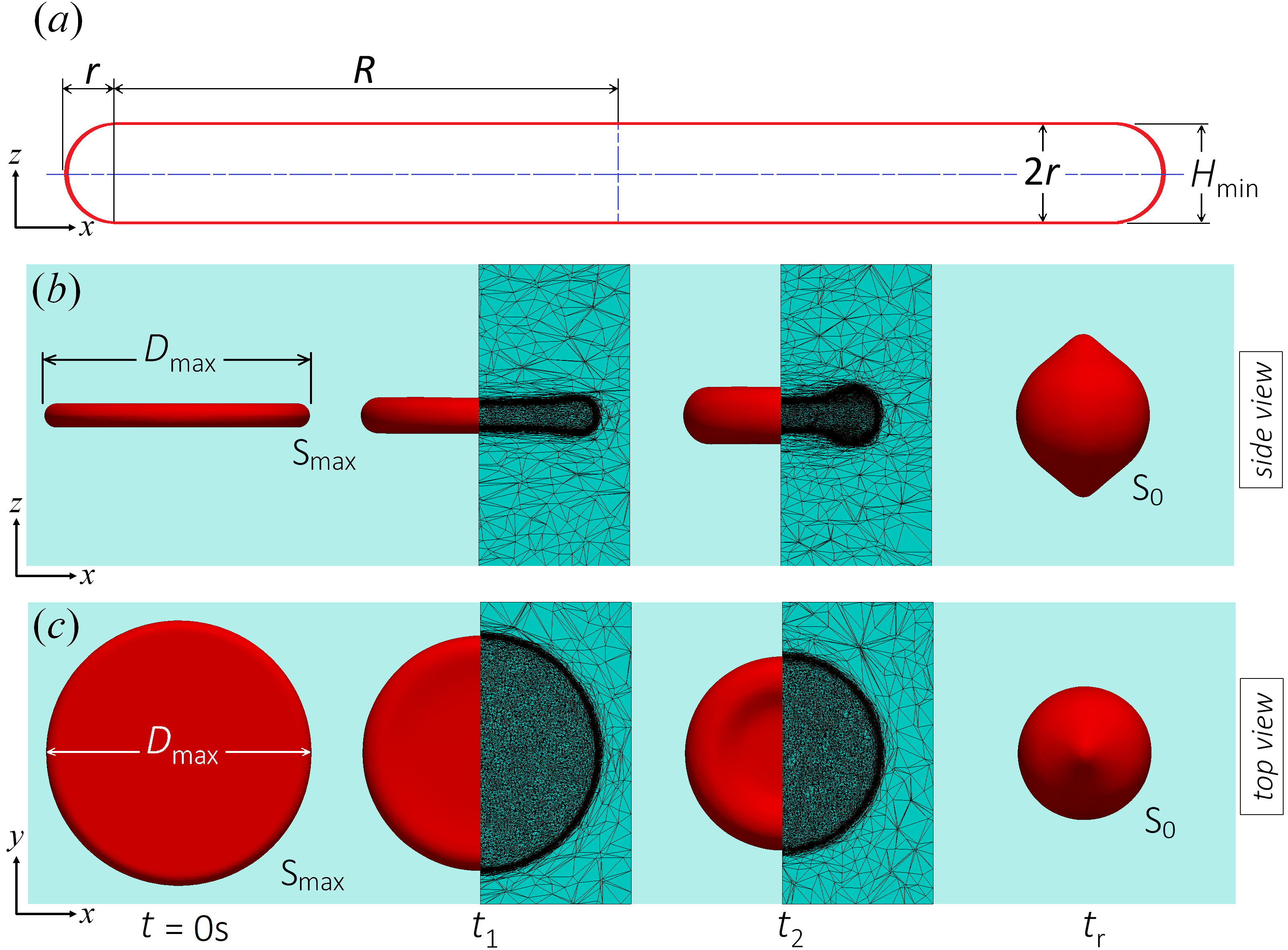}
\vspace*{-0.6cm}
\caption{(\textit{a}) Schematic of a drop morphology (i.e. pancake-like) at $t=0$s (side view). (\textit{b}) Side view and (\textit{c}) top view snapshots of a multiphase numerical simulation: typical retraction of a levitating Newtonian drop (in red) of initial diameter $D_{max}$, surface area $S_{max}$, initial thickness $H_{min}$, density $\rho$, viscosity $\eta$, and surface tension $\sigma$. The drop retracts until its (quasi)-spherical minimum surface $S_0$ at time $t_r$. Both the drop and the solid surface are surrounded by air (in blue). The mesh (depicted with black lines) is composed of approximately 10$^6$ elements whose minimum size is 1$\mu$m. A movie showing a typical numerical simulation is available online in which $D_{max} =$ 22mm, $H_{min} =$ 2mm, $D_0 =$ 11.15mm, $\eta =$ 0.001089Pa$\cdot$s, $\rho =$ 100kg/m$^3$, $\sigma =$ 0.2N/m (https://youtu.be/hoV9A2PT4jE).}
\label{fig-2}
\end{figure*}

As previously highlighted, we present here a theoretical and numerical study on the retraction of levitating Newtonian drops exhibiting the pancake-like shape illustrated in figure \ref{fig-2}(\textit{a}) at the initial instant $t = 0$s. Hence, at the initial instant of the retraction process, the considered drops present a maximum diameter $D_{max}= 2\left( R+r \right)$, minimum height $H_{min} = 2r$, surface $S_{max} = 2\pi R^2 + 2 \pi R \pi r$, density $\rho$, viscosity $\eta$, and surface tension $\sigma$, levitating under zero-gravity conditions with zero velocity. They are surrounded by air with density $\rho_{air}$ and viscosity $\eta_{air}$. Furthermore, they retract under capillary pressure effects, gradually assuming a surface $S_0 = 4 \pi R_0^2$ at instant $t_r$. As underlined by figure \ref{fig-1}, low viscosity drops tend to oscillate before their kinetic energy is completely dissipated (sub-figures \ref{fig-1}\textit{a}-\textit{c}). Consequently, when their surface becomes equal to $S_0 = 4 \pi R_0^2$ [where $ 2 R_0 = D_0$ and $R_0 = [(3/2)R^2 r + (3/4)\pi R r^2]^{1/3}$ by volume conservation] for the first time (at $t = t_r$), they present a non-zero velocity, which ultimately gives them two necks, as illustrated in sub-figures \ref{fig-1}(\textit{a}) and \ref{fig-2}(\textit{b})-(\textit{c}) with side and top views, respectively. In contrast, high viscosity drops tend to exhibit a zero-velocity spherical shape at $t_r$, as illustrated by sub-figure \ref{fig-1}(\textit{d}). These different retraction dynamics will be further discussed in section \ref{RD}. 

Our computational method is based on a massively parallel finite element library \citep[CIMLIB-CFD;][]{Coupez-13, Hachem-16} devoted to non-Newtonian multiphase flows \citep{Riber-16, Valette-19, Pereira-19, Pereira-20, Valette-21, Isukwem-24a, Isukwem-24b, Isukwem-24d, Isukwem-24e}. More specifically, we apply the momentum conservation equation presented below to the solenoidal flow ($\boldsymbol{\nabla \cdot u} = 0$) described earlier (see figure \ref{fig-2}):     
\begin{equation}
\rho \left( \frac{\partial \boldsymbol{u}}{\partial t} + \boldsymbol{u} \cdot \nabla \boldsymbol{u} - \boldsymbol{g} \right) = - \nabla p + \nabla \cdot \boldsymbol{\tau} + \boldsymbol{f_{st}} \, ,
\label{eq:cons-mom}
\end{equation} 
in which $\boldsymbol{u}$, $p$, $\boldsymbol{\tau}$, $\nabla$, $\boldsymbol{g}$, $\nabla \cdot$ and $\boldsymbol{f_{st}}$ are, respectively, the velocity vector  $\boldsymbol{u} = \left\lbrace u_x, u_y, u_z \right\rbrace  $, the pressure, the extra-stress tensor, the gradient operator, the gravity vector, the divergence operator, and a capillary term related to the surface tension force. The latter is defined as $\boldsymbol{f_{st}} = -\sigma \kappa \Phi \boldsymbol{n}$, where $\sigma$, $\kappa$, $\Phi$, and $\boldsymbol{n}$ are the drop surface tension with the surrounding gas, the curvature of the drop-gas interface, the Dirac function locating the drop-gas interface, and its normal vector, respectively. In addition, the extra-stress tensor is given by $\boldsymbol{\tau} = \eta \boldsymbol{\dot{\gamma}}$, in which $\boldsymbol{\dot{\gamma}}$ represents the \textit{rate-of-strain} tensor defined as $\boldsymbol{\dot{\gamma}} = \left( \boldsymbol{\nabla u} + \boldsymbol{\nabla u}^T \right)$. The norm of $\boldsymbol{\dot{\gamma}}$ is called \textit{deformation rate}, being defined as $|| \boldsymbol{\dot{\gamma}}|| = \left( \frac{1}{2} \boldsymbol{\dot{\gamma}}: \boldsymbol{\dot{\gamma}} \right) ^{\frac{1}{2}}$. 

\begin{figure}%[h!]
\centering
\includegraphics[angle=0, scale=0.25]{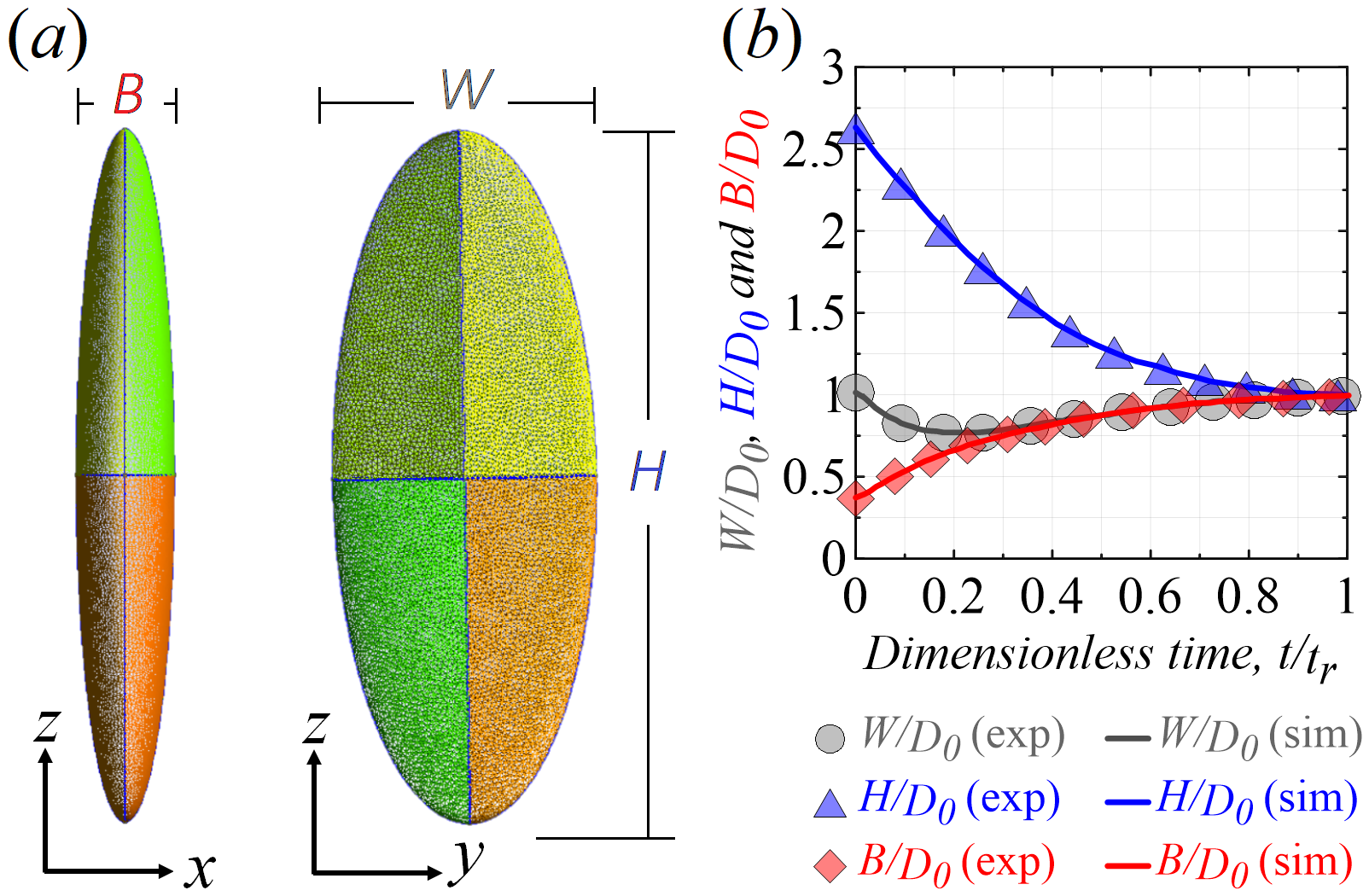}
\vspace*{-0.1cm}
\caption{Retraction dynamics of a prolate polyurethane drop (viscosity = 0.485Pa$\cdot$s) dispersed in a viscous polydimethylsiloxane (PDMS) continuous phase (viscosity = 48.5Pa$\cdot$s). 
The surface tension between the phases is 0.0216N/m. (\textit{a}) Drop morphology at $t =$ 0s. (\textit{b}) Retraction processes. The symbols indicate the experimental data reported by Assighaou and Benyahia \citep{Assighaou-10}: length $L$ (blue triangles), width $W$ (grey circles) and height $H$ (red diamonds). Our numerical results are represented by the lines: length $L$ (blue solid line), width $W$ (grey dashed line) and height $H$ (red dash-dotted line). These quantities are made dimensionless by the diameter of the equilibrium spherical drop ($D_0$ = 500$\mu$m) and plotted as a function of the dimensionless time $t/t_r$.}    
\vspace*{-0.2cm}
\label{fig-validation}
\end{figure}

The numerical methods used here are based on a \textit{Variational Multiscale Method} (VMS) coupled with an anisotropic mesh adaptation method \citep{Valette-19, Pereira-19, Pereira-20, Valette-21, Isukwem-24a, Isukwem-24b, Isukwem-24d, Isukwem-24e}. The meshes used are composed of approximately 10$^6$ elements whose minimum size is 1$\mu$m (see the black lines in sub-figures \ref{fig-2}\textit{c}-\textit{d}). The evolution of the interface of the 3D objects over time is described using a Level-Set function \citep{Coupez-13, Hachem-16}. It is important to emphasise that our numerical multiphase framework has been validated for a variety of flow scenarios, which includes the impact of drops \citep{Isukwem-24a, Isukwem-24b, Isukwem-24d, Isukwem-24e}, dam-breaks \citep{Valette-21}, the stretching and breakup of filaments \citep{Valette-19}, and the development of buckling instabilities \citep{Pereira-19, Pereira-20} using Newtonian and non-Newtonian fluids. Moreover, a numerical validation based on experimental-numerical comparisons devoted to the retraction of a viscous prolate drop \citep{Assighaou-10} is provided by figure \ref{fig-validation}. The length $L$, width $W$ and height $H$ of the ellipsoid (made dimensionless by $D_0$) suspended in fluid 100 times more viscous are plotted as a function of the dimensionless retraction time $t/t_r$. Clearly, the numerical results are in line with the experiments. The good agreement between simulations and experiments underlined throughout the numerous studies mentioned above gives us the confidence to explore numerically the physical mechanisms driving the retraction of levitating drops.
\begin{figure*}%[h!]
\centering
\includegraphics[angle=0, scale=0.42]{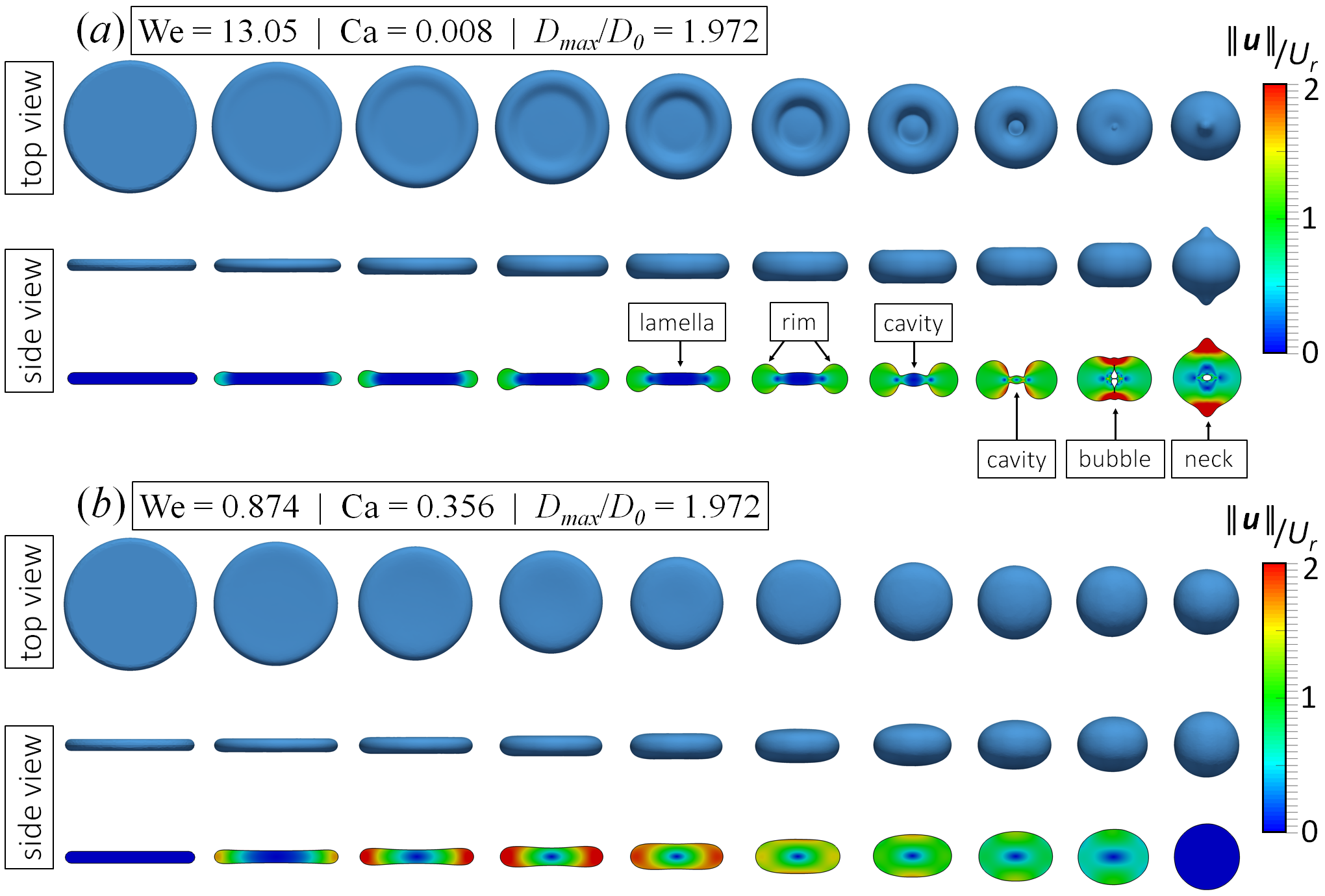}
\vspace*{-0.4cm}
\caption{Retraction dynamics of a levitating Newtonian drop with $D_{max}/D_0 = 1.972$ (pancake-like), for two different $\mathrm{We}$-$\mathrm{Ca}$ couples: (\textit{a}) $\mathrm{We} = 13.05$, $\mathrm{Ca} = 0.008$ ($\Pi_{\mathrm{We}} = 0.978$ and $\Pi_{\mathrm{Ca}} = 0.022$) and (\textit{b}) $\mathrm{We} = 0.874$, $\mathrm{Ca} = 0.356$ ($\Pi_{\mathrm{We}} = 0.066$ and $\Pi_{\mathrm{Ca}} = 0.934$). Each sub-figure is composed of ten images illustrating the retraction of the drop at ten different times ($t=0$ to $t=t_r$), i.e., from maximum spread ($S_{max}$) to complete retraction ($S_0$). The interval $\Delta t$ between two snapshots is $t_r/9$ms. The top row and middle row images show the instantaneous morphology of the drops from the top view and side view respectively, while the bottom row images show the contours of the dimensionless instantaneous velocity $||\boldsymbol{u}||/U_r$ from the side view.}
\label{fig-3}
\end{figure*}

In the present work, the retracting objects are micrometric/millimetric/centimetric. Their initial maximum spreading $D_{max}$ varies between 1mm and 50mm, while their diameter $D_0$ at the end of the retraction process is between 0.1mm and 12mm ($1 < D_{max}/D_0 \leq 10$). Note that by imposing $D_{max}$ and $D_0$, one automatically imposes $R$ and $r$ by volume conservation. A wide range of drop viscosity, density and surface tension is considered: 0.0005Pa$\cdot$s $\leq \eta \leq$ 10Pa$\cdot$s; 10kg/m$^3 \leq \rho \leq$ 5000kg/m$^3$; and 0.001N/m $ \leq \sigma \leq $ 0.5N/m. In addition, the viscosity and density of the air are fixed at $\eta_{air} =$ 10$^{-5}$Pa$\cdot$s and  $\rho_{air} =$ 1kg/m$^3$. Under zero gravity conditions ($||\boldsymbol{g}|| =$ 0m/s$^2$), the mentioned length and properties lead to retraction times $t_r$ typically varying in between 1ms (for low viscosity drops) and 100ms (for high viscosity drops), in terms of order of magnitude. 

The drops are contained in a computational cube (i.e. computational domain; not shown in figure \ref{fig-2} for brevity) whose dimensions are 100 times larger than $D_{max}$ (which completely mitigates wall effects on the drop dynamics). No-slip conditions are imposed at its lateral walls ($x$ and $y$ directions), while zero normal stresses are imposed at the top and bottom walls ($z$ direction). A movie showing a typical numerical simulation is available online in which $D_{max} =$ 22mm, $H_{min} =$ 2mm, $D_0 =$ 11.15mm, $\eta =$ 0.001089Pa$\cdot$s, $\rho =$ 100kg/m$^3$, $\sigma =$ 0.2N/m (https://youtu.be/hoV9A2PT4jE).

The dimensionless numbers dominating the problem $\Pi_i$ are obtained from the Buckingham-$\Pi$ theorem \citep{Buckingham-14, Buckingham-15} whose variables are $D_{max}$, $D_0$, $\eta$, $\rho$, $\sigma$, and $U_r$, in which the latter denotes a characteristic retraction velocity (we will give more details on this term in the following lines). The fundamental units are mass [kg], distance [m] and time [s]. This results in three important dimensionless quantities:    
\begin{equation}
\Pi_1 = \frac{D_{max}}{D_0} \, ,
\label{eq:pi-1}
\end{equation}
\begin{equation}
\Pi_2 = \frac{\rho U_r^2}{\sigma/D_0} \, ,
\label{eq:pi-2}
\end{equation}  
\begin{equation}
\Pi_3 = \frac{\eta(U_r/D_0)}{\sigma/D_0} \, ,
\label{eq:pi-3}
\end{equation}  
where $\Pi_2 = \mathrm{We}$ ($\mathrm{We}$ is the Weber number) and $\Pi_3= \mathrm{Ca}$ ($\mathrm{Ca}$ denotes the capillary number). The effects of these three dimensionless numbers on the drop retraction process are highlighted in section \ref{RD}. In other words, our results are presented in terms of these three parameters.

Lastly, a characteristic retraction velocity $U_r$ can be defined by assuming that the surface energy of the pancake-like drop [$\sigma (2 \pi R^2 + 2 \pi R \pi r)$] at $t = 0$s is converted into spherical-based surface energy ($\sigma 4 \pi R_0^2$), kinetic energy [$(\rho U_r^2 /2)(4 \pi R_0^3 /3)$] and dissipated energy [$3 \eta U_r  \pi (R+r)^2$] at $t = t_r$ (the presence of the prefactor 3 in the dissipation term is related to the uniaxial extensional nature of the viscous-dominated flow, which will be clarified in section \ref{RD}). This leads to the following equation     
\begin{equation}
\frac{1}{6}\rho U_r^2  + \frac{3}{2} \frac{\eta}{2R_0} \frac{(R+r)^2}{R_0^2} U_r -  \frac{\sigma}{2R_0} \left(   \frac{R^2}{R_0^2}   +  \frac{\pi R r}{R_0^2}    -  2   \right)  = 0   \, ,
\label{eq:ur-1}
\end{equation}  
which can be solved for $U_r$
\begin{equation}
U_r = \frac{ - \frac{3\eta (R+r)^2}{4R_0^3}   +   \sqrt{  \left[ \frac{3\eta (R+r)^2}{4R_0^3}   \right]^2  +  \frac{\rho \sigma}{3 R_0}  \left(  \frac{R^2}{R_0^2} + \frac{\pi R r}{R_0^2}  - 2 \right)  }   }{\frac{\rho}{3}}         \, .
\label{eq:ur-2}
\end{equation} 
This equation is used in the calculation of $\mathrm{We}$ and $\mathrm{Ca}$. Additionally, by re-arranging equation \ref{eq:ur-1}, we find that  
\begin{equation}
1 = \frac{\mathrm{We}}{6\left(  \frac{R^2}{R_0^2} + \frac{\pi R r}{R_0^2} - 2  \right) } + \frac{3\mathrm{Ca}(R+r)^2 }{2 R_0^2\left(  \frac{R^2}{R_0^2} + \frac{\pi R r}{R_0^2} - 2  \right) }       \, ,
\label{eq:pi-extra}
\end{equation}
where the first right-hand side term will be referred to as $\Pi_{\mathrm{We}}$, and the second one will be denoted as $\Pi_{\mathrm{Ca}}$. These two dimensionless numbers express aspect ratio, Weber number and capillary number effects. Consequently, they can also be used to depict our results. For this reason, in section \ref{RD}, we give their values, in addition to $D_{max}/D_0$, $\mathrm{We}$ and $\mathrm{Ca}$. Finally, it is worth noting that $\Pi_{\mathrm{We}}$ and $\Pi_{\mathrm{Ca}}$ are bounded between 0 and 1. In inviscid scenarios, $\Pi_{\mathrm{We}} = 1$ and $\Pi_{\mathrm{Ca}} = 0$. On the other hand, for inertia-free flow cases, $\Pi_{\mathrm{We}} = 0$ and $\Pi_{\mathrm{Ca}} = 1$. As indicated by equation \ref{eq:pi-extra}, $\Pi_{\mathrm{We}} + \Pi_{\mathrm{Ca}} = 1$.      

\begin{figure*}%[h!]
\centering
\includegraphics[angle=0, scale=0.234]{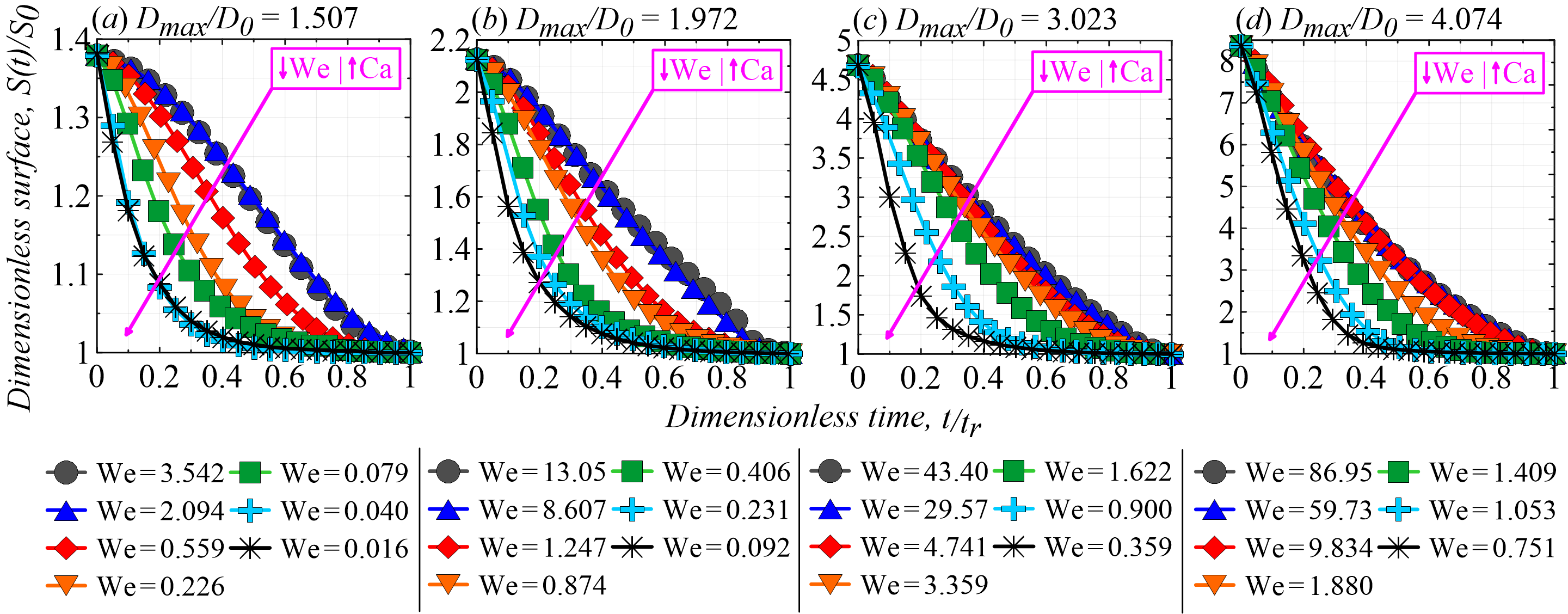}
\vspace*{-0.6cm}
\caption{Instantaneous dimensionless surface $S(t)/S_0$ as a function of the dimensionless time $t/t_r$ for four different pancake drops: (\textit{a}) $D_{max} = 1.507 D_0$, (\textit{b}) $D_{max} = 1.972 D_0$, (\textit{c}) $D_{max} = 3.023 D_0$ and (\textit{d}) $D_{max} = 4.074 D_0$. For each displayed case, seven different $\mathrm{We}$-$\mathrm{Ca}$ couples are considered. Each $\mathrm{We}$-$\mathrm{Ca}$ is represented by a specific symbol. Decreasing $\mathrm{We}$ when moving downwards on the plots implies increasing $\mathrm{Ca}$, as indicated by equation \ref{eq:pi-extra}.}
\label{fig-4}
\end{figure*}

%%%%%%%%%%%%%%%%%%%%%%%%%%%%%%%%%%%%%%%%%%%%%%%%%%%%%%%%%%%%%%%%%%%%%%%%%%%%%%
%%%%%%%%%%%%%%%%%%%%%%%%%%%%%%%%%%%%%%%%%%%%%%%%%%%%%%%%%%%%%%%%%%%%%%%%%%%%%%
\section{Results and Discussion} \label{RD}
%%%%%%%%%%%%%%%%%%%%%%%%%%%%%%%%%%%%%%%%%%%%%%%%%%%%%%%%%%%%%%%%%%%%%%%%%%%%%%
%%%%%%%%%%%%%%%%%%%%%%%%%%%%%%%%%%%%%%%%%%%%%%%%%%%%%%%%%%%%%%%%%%%%%%%%%%%%%%
%\vspace*{-0.5cm}

In sub-figures \ref{fig-3}(\textit{a})-(\textit{b}), snapshots illustrate the retraction of drops with $D_{max}/D_0$ = 1.972 at two $\mathrm{We}$-$\mathrm{Ca}$ couples, respectively: $\mathrm{We} =$ 13.05 and $\mathrm{Ca} =$ 0.008 ($\Pi_{\mathrm{We}} = $0.978 and $\Pi_{\mathrm{Ca}} = $0.022) and $\mathrm{We} =$ 0.874 and $\mathrm{Ca} =$ 0.356 ($\Pi_{\mathrm{We}} = $0.066 and $\Pi_{\mathrm{Ca}} = $0.934). The interval $\Delta t$ between two subsequent images corresponds to 1/9 of the necessary time to achieve $t_r$ ($\Delta t = t_r/9$). Views of the top (upper line) and side (middle and bottom lines) are provided. Additionally, contours of the norm of the instantaneous velocity $||\boldsymbol{u}(x,y,z,t)||$ (made dimensionless by the characteristic retraction velocity $U_r$) on the centre $x$-$z$ plane are displayed (bottom line). Different retraction dynamics are highlighted.

In the flow case depicted in sub-figure \ref{fig-3}(\textit{a}), a rim is formed at the beginning of the retraction process, which, combined with the diameter decrease and the thinning of the lamellar part of the drop, leads to the development of an air-filled cavity at later instants ($0.5 t_r < t < 0.8 t_r$). Nevertheless, the cavity collapses, entrapping an air bubble and triggering an axial movement leading to the formation of a liquid neck ($0.8 t_r < t < t_r$) and ultimately a jet \citep[$t > t_r$; not shown for brevity;][]{Worthington-97, Worthington-00, Eggers-08a, Gekle-10a, Gekle-10b}. Interestingly, the decrease of $\mathrm{We}$ (e.g., the increase of $\mathrm{Ca}$) leads to a more homogeneous retraction, as displayed by sub-figure \ref{fig-3}(\textit{b}). From the beginning of the process, the diameter reduction of the drop is accompanied by the increase of its height, thanks to which a zero-velocity spherical shape is found at $t = t_r$. From this point on, drop movement ceases (similar to the high viscosity drop illustrated in sub-figures \ref{fig-1}\textit{d}-\textit{f}).  

\begin{figure*}%[h!]
\centering
\includegraphics[angle=0, scale=0.215]{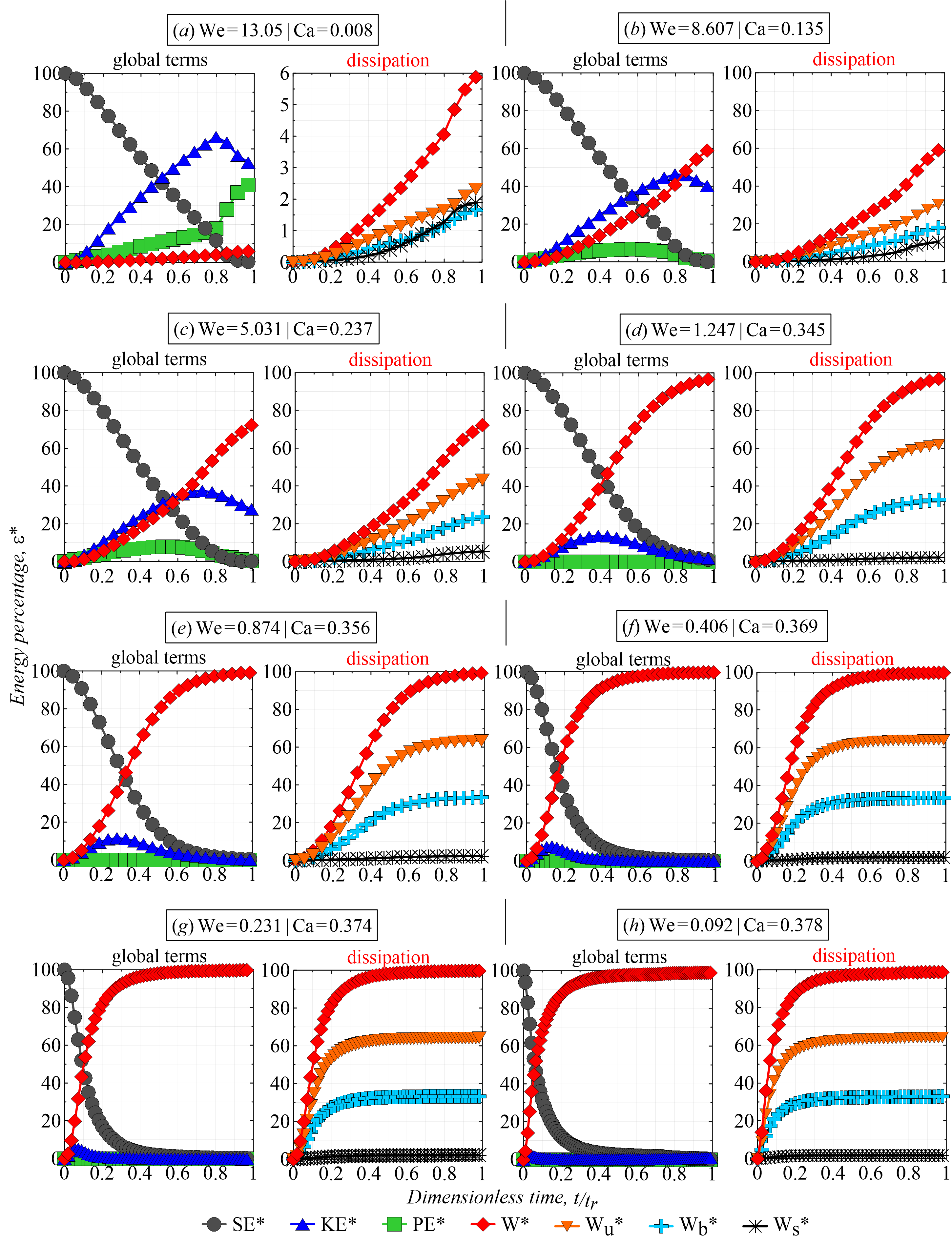}
\vspace*{-0.4cm}
\caption{Energy percentage $\epsilon^{\ast}$ during the retraction of levitating Newtonian drops with $D_{max}/D_0 = 1.972$ for eight $\mathrm{We}$-$\mathrm{Ca}$) couples: (\textit{a}) $\mathrm{We} = 13.05$, $\mathrm{Ca} = 0.008$ ($\Pi_{\mathrm{We}} = 0.978$ and $\Pi_{\mathrm{Ca}} = 0.022$); (\textit{b}) $\mathrm{We}$ = 8.607, $\mathrm{Ca}$ = 0.135 ($\Pi_{\mathrm{We}} = 0.645$ and $\Pi_{\mathrm{Ca}} = 0.355$); (\textit{c}) $\mathrm{We}$ = 5.031, $\mathrm{Ca}$ = 0.237 ($\Pi_{\mathrm{We}} = 0.377$ and $\Pi_{\mathrm{Ca}} = 0.623$); (\textit{d}) $\mathrm{We}$ = 1.247, $\mathrm{Ca}$ = 0.345 ($\Pi_{\mathrm{We}} = 0.093$ and $\Pi_{\mathrm{Ca}} = 0.906$); (\textit{e}) $\mathrm{We}$ = 0.874, $\mathrm{Ca}$ = 0.356 ($\Pi_{\mathrm{We}} = 0.065$ and $\Pi_{\mathrm{Ca}} = 0.934$); (\textit{f}) $\mathrm{We}$ = 0.406, $\mathrm{Ca}$ = 0.369 ($\Pi_{\mathrm{We}} = 0.030$ and $\Pi_{\mathrm{Ca}} = 0.969$); (\textit{g}) $\mathrm{We}$ = 0.231, $\mathrm{Ca}$ = 0.374 ($\Pi_{\mathrm{We}} = 0.017$ and $\Pi_{\mathrm{Ca}} = 0.983$); (\textit{h}) $\mathrm{We}$ = 0.092, $\mathrm{Ca}$ = 0.378 ($\Pi_{\mathrm{We}} = 0.007$ and $\Pi_{\mathrm{Ca}} = 0.993$).} 
\label{fig-5}
\end{figure*}

Further retraction dynamics analyses are presented by figure \ref{fig-4}, in which the instantaneous dimensionless surface $S(t)/S_0$ is plotted as a function of the dimensionless time $t/t_r$. Effects of $\mathrm{We}$ and $\mathrm{Ca}$ on $S(t)/S_0$ are shown for four $D_{max}/D_0$ values: (\textit{a}) 1.507; (\textit{b}) 1.972; (\textit{c}) 3.023; and (\textit{d}) 4.074. Each curve represents a flow case at a specific $\mathrm{We}$-$\mathrm{Ca}$ couple. Clearly, $S(t)/S_0$ profiles drastically change when decreasing $\mathrm{We}$ (and consequently increasing $\mathrm{Ca}$ for a fixed $D_{max}/D_0$, as pointed out by equation \ref{eq:pi-extra}), as underlined by the magenta arrows. High $\mathrm{We}$ flow cases exhibit a quasi-linear surface decay (see, for instance, at $\mathrm{We} =$ 12.76, in sub-figure \ref{fig-4}\textit{b}) for which 
\begin{equation}
\frac{(R + r - R_0)}{t_r} \approx U_r  ~~~~ (\mathrm{similar~to~figure~\ref{fig-1}\textit{c}})        \, ,
\label{eq:linear-decay}
\end{equation}
whereas low $\mathrm{We}$ cases present an exponential $S(t)/S_0$ profile, with an analogous exponential height increase 
\begin{equation}
\frac{H}{2R_0} \approx \left( \frac{r}{R_0} - 1  \right)  e^{-k ~ ln(R_0/r) ~ t/t_r}  +  1  ~~~ (\mathrm{see~figure~\ref{fig-1}\textit{f}})        \, ,
\label{eq:exponential-decay}
\end{equation}
in which $k$ is a prefactor and $(k/t_r)\ln(R_0/r)$ denotes a characteristic deformation rate ${\dot{\gamma}}_r$. These two distinct profiles indicate the existence of different retraction regimes, as discussed in the following lines.   

The physical mechanisms driving the retraction dynamics are investigated in light of energy analyses. In figure \ref{fig-5}, energy percentage curves are plotted as a function of the dimensionless time: relative surface energy $SE$, kinetic energy $KE$, dissipated energy $W$, and pressure energy $PE$. These terms are respectively defined as  
\begin{equation}
SE = \int_{S}^{} \sigma dS - \sigma 4\pi R_0^2    \, ,
\label{eq:SE}
\end{equation}
\begin{equation}
KE = \int_{V}^{}\frac{1}{2} \rho {\parallel\boldsymbol{u}\parallel}^2 dV    \, ,
\label{eq:KE}
\end{equation}
\begin{equation}
W = \int_{t}^{} \int_{V}^{} \frac{\eta}{2} {\parallel \boldsymbol{{\dot{\gamma}}}  \parallel}^{2} dV dt    \, ,
\label{eq:W}
\end{equation}
\begin{equation}
PE = \int_{t}^{} \int_{V}^{} (\nabla p) \cdot \boldsymbol{u} dV dt    \, ,
\label{eq:PE}
\end{equation}
where $\sigma 4\pi R_0^2$ represent the surface energy of the drop at the equilibrium, $V$ is the drop volume, and $S$ is the instantaneous drop surface \citep[silimar to][]{Valette-21, Sanjay-21, Isukwem-24a, Isukwem-24b}. Furthermore, the dissipated energy is split into the uniaxial $W_u = \int_{t}^{} \int_{V}^{} \frac{\eta}{2}  \frac{\dot{\gamma}_{zz}^{2}}{2} dV dt$, biaxial deformation contribution $W_b = \int_{t}^{} \int_{V}^{} \frac{\eta}{2} \left( \frac{\dot{\gamma}_{xx}^{2} + \dot{\gamma}_{yy}^{2}}{2} \right)  dV dt$, and its complementary shear-based part $W_s = W - (W_u + W_b)$ (where $\dot{\gamma}_{xx} = 2\partial{u_x}/\partial{x}$, $\dot{\gamma}_{yy} = 2\partial{u_y}/\partial{y}$, and $\dot{\gamma}_{zz} = 2\partial{u_z}/\partial{z}$). Each energy term is made dimensionless by the total energy of the system $E_0 = SE(t=0)$: $SE^{\ast} = SE/E_0 ~ \times$ 100[\%] (grey circles), $KE^{\ast} = KE/E_0 ~ \times$ 100[\%] (blue triangles), $W^{\ast} = W/E_0 ~ \times$ 100[\%] (red diamonds), $W_u^{\ast} = W_u/E_0 ~ \times$ 100[\%] (inverted orange triangles), $W_b^{\ast} = W_b/E_0 ~ \times$ 100[\%] (blue crosses), and $W_s^{\ast} = W_s/E_0 ~ \times$ 100[\%] (black asterisks). In addition, each sub-figure is related to a specific $\mathrm{We}$-$\mathrm{Ca}$ couple at $D_{max}/D_0 =$ 1.972:
(\textit{a}) $\mathrm{We} =$ 13.05 and $\mathrm{Ca} =$ 0.008 ($\Pi_{\mathrm{We}} =$ 0.978 and $\Pi_{\mathrm{Ca}} =$ 0.022); (\textit{b}) $\mathrm{We} =$ 8.607 and $\mathrm{Ca} =$ 0.135 ($\Pi_{\mathrm{We}} =$ 0.645 and $\Pi_{\mathrm{Ca}} =$ 0.355); (\textit{c}) $\mathrm{We} =$ 5.031 and $\mathrm{Ca} =$ 0.237 ($\Pi_{\mathrm{We}} =$  0.377 and $\Pi_{\mathrm{Ca}} =$ 0.623); (\textit{d}) $\mathrm{We} =$ 1.247 and $\mathrm{Ca} =$ 0.345 ($\Pi_{\mathrm{We}} =$ 0.093 and $\Pi_{\mathrm{Ca}} =$ 0.906); (\textit{e}) $\mathrm{We} =$ 0.874 and $\mathrm{Ca} =$ 0.356 ($\Pi_{\mathrm{We}} =$ 0.065 and $\Pi_{\mathrm{Ca}} =$ 0.934); (\textit{f}) $\mathrm{We} =$ 0.406 and $\mathrm{Ca} =$ 0.369 ($\Pi_{\mathrm{We}} =$ 0.030 and $\Pi_{\mathrm{Ca}} =$ 0.969); (\textit{g}) $\mathrm{We} =$ 0.231 and $\mathrm{Ca} =$ 0.374 ($\Pi_{\mathrm{We}} =$ 0.017 and $\Pi_{\mathrm{Ca}} =$ 0.983); and (\textit{h}) $\mathrm{We} =$ 0.092 and $\mathrm{Ca} =$ 0.378 ($\Pi_{\mathrm{We}} =$ 0,007 and $\Pi_{\mathrm{Ca}} =$ 0.993). 

For all cases shown in figure \ref{fig-5}, the retraction is induced by the initial surface energy of the drop, $SE^{\ast}(t = 0) =$ 100\%, which is later on converted into kinetic energy and dissipated by viscous effects. At $\mathrm{We} =$ 13.05 and $\mathrm{Ca} =$ 0.008 ($\Pi_{\mathrm{We}} =$ 0.978 and $\Pi_{\mathrm{Ca}} =$ 0.022; sub-figure \ref{fig-5}\textit{a}), $SE^{\ast}$ is primarily converted into $KE^{\ast}$, indicating that the retraction dynamics is dominated by the balance between capillary and inertial stresses. This is a capillary-inertial regime. As $\mathrm{We}$ decreases (and consequently $\mathrm{Ca}$ increases), dissipation becomes more pronounced (sub-figures \ref{fig-5}\textit{b}-\textit{c}), eventually dominating the retraction (sub-figures \ref{fig-5}\textit{d}-\textit{h}). The drop is driven by a competition between capillary and viscous stresses in such a scenario. This is a capillary-viscous regime. Lastly, pressure energy remains marginal for the mentioned regimes.     

Focusing on the dissipation terms, it is crucial to observe that $W_u^{\ast}$, $W_b^{\ast}$ and $W_s^{\ast}$ are comparable to each other in capillary-inertial flow scenarios. However, as dissipations become dominating, the flow becomes essentially linked to $\dot{\gamma}_{xx}$, $\dot{\gamma}_{yy}$ and $\dot{\gamma}_{zz}$, exhibiting a pronounced uniaxial character (in sub-figure \ref{fig-5}\textit{h}, for instance, $W_u^{\ast} =$ 65\%, $W_b^{\ast} =$ 35\% and $W_s^{\ast} =$ 0\% when $SE^{\ast} = KE^{\ast} =$ 0\%; $ \frac{1}{2} ||\dot{\gamma}_{xx}|| \approx  \frac{1}{2} ||\dot{\gamma}_{yy}|| \approx ||\dot{\gamma}_{zz}||$). This result, together with the axial movement of the drop observed throughout the retraction process illustrated by sub-figure \ref{fig-3}(\textit{b}) and the exponential $S(t)/S_0$ and $H(t)/D_0$ profiles highlighted by sub-figure \ref{fig-1}(\textit{f}), reveals a steady flow with a constant deformation rate proportional to $(1/t_r)\ln(R_0/r)$ \citep{Anna-01}, as detailed along the following lines.

The behaviour stressed above is observed for all $D_{max}/D_0$ considered here (not shown for brevity). Furthermore, it is worth noting that $\Pi_{\mathrm{We}} \geq$ 0.9 and $\Pi_{\mathrm{Ca}} \leq$ 0.1 for flow cases dominated by a capillary-inertial balance (sub-figure \ref{fig-5}\textit{a}). On the other hand, for capillary-viscous-driven scenarios, $\Pi_{\mathrm{We}} \leq$ 0.1 and $\Pi_{\mathrm{Ca}} \geq$ 0.9 (sub-figures \ref{fig-5}\textit{d}-\textit{h}). 

\begin{figure}%[h!]
\centering
\includegraphics[angle=0, scale=0.3]{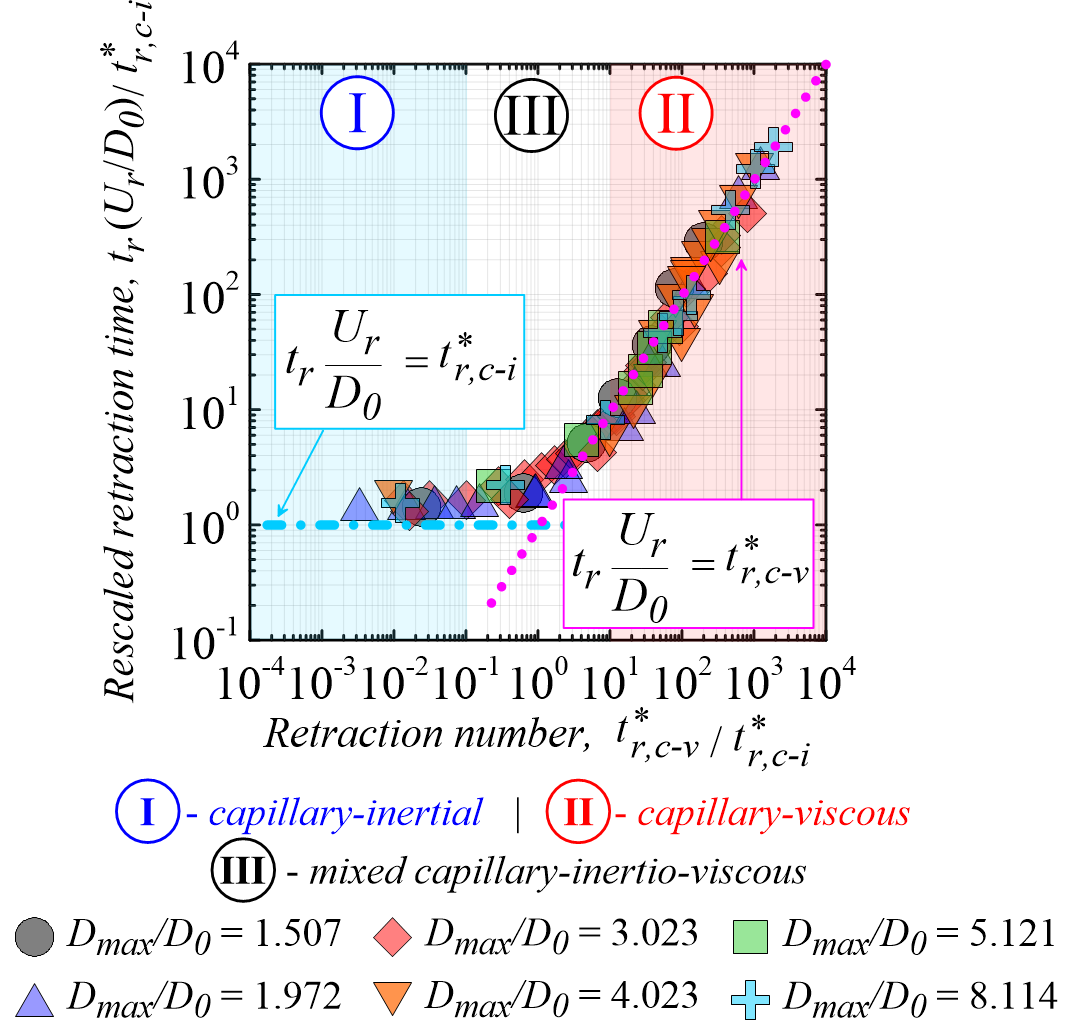}
\vspace*{-0.25cm}
\caption{Rescaled retraction time $t_r (U_r/D_0)/t_{r,c-i}^*$ as a function of the retraction number ${t_{r,c-v}^*}/{t_{r,c-i}^*}$. Simulations for six stretched pancake-like Newtonian drops (grey circles for $D_{max}/D_0 = 1.507$, blue triangles for $D_{max}/D_0 = 1.972$, red diamonds for $D_{max}/D_0 = 3.023$, orange inverted triangles for $D_{max}/D_0 = 4.023$, green squares for $D_{max}/D_0 = 5.121$ and sky blue crosses for $D_{max}/D_0 = 8.114$). They collapse across one path divided into three regions linked to different retraction regimes: capillary-inertial (I; blue box), for which $t_r (U_r/2R_0 ) = t_{r,c-i}^*$ (blue dash-dotted line); the capillary-viscous (II; red box) where $t_r (U_r/2R_0 ) = t_{r,c-v}^*$ (magenta dotted line); and a mixed regime (III; white box), for which capillary, inertial , and viscous effects all play an important role. As underlined by our theoretical analyses, a mixed regime emerges when $0.1 \leq {t_{r,c-v}^*}/{t_{r,c-i}^*} \leq 10$.}
\label{fig-6}
\vspace*{-0.3cm}
\end{figure}

The energy budgets illustrated in figure \ref{fig-5} highlight the existence of at least three retraction regimes: the capillary-inertial regime (sub-figure \ref{fig-5}\textit{a}), the mixed capillary-inertio-viscous regime (sub-figures \ref{fig-5}\textit{b}-\textit{c}), and the capillary-viscous regime (sub-figures \ref{fig-5}\textit{d}-\textit{h}). For the capillary-inertial regime, the difference of surface energy of the drop at $t/t_r = 0$ and $t/t_r = 1$ [$\sigma (2\pi R^2 + 2 \pi R \pi r-4 \pi R_0^2)$, where $(2 \pi R^2 + 2 \pi R \pi r)$ is the surface of the drop at $t/t_r = 0$ and $4 \pi R_0^2$ represents the surface at $t/t_r = 1$] is converted into kinetic [$(\rho U_r^2/2)(4 \pi R_0^3/3)$]. By equating these energy terms, we find 
\begin{equation}
1 = \frac{\mathrm{We}}{6\left(  \frac{R^2}{R_0^2} + \frac{\pi R r}{R_0^2} - 2  \right) }  \, ,
\label{eq:cap-in-1}
\end{equation}
which corresponds to the first right-hand term of equation \ref{eq:pi-extra}. In addition, as pointed out by figures \ref{fig-1} and \ref{fig-4}, and equation \ref{eq:linear-decay}, in the capillary-inertial regime $t_r \approx (R + r - R_0)/U_r$ and consequently 
\begin{equation}
t_r \left(  \frac{U_r}{2R_0}  \right) \approx  \frac{R + r - R_0}{2R_0}   \, .
\label{eq:cap-in-2}
\end{equation}
By multiplying equations \ref{eq:cap-in-1} and \ref{eq:cap-in-2}, we find
\begin{equation}
t_r \left(  \frac{U_r}{2R_0}  \right) \approx  \mathrm{We} \frac{R+r-R_0}{ 12 R_0 \left(  \frac{R^2}{R_0^2} + \frac{\pi R r}{R_0^2} - 2  \right)  }  =  t_{r,c-i}^{\ast}  \, ,
\label{eq:cap-in-scaling}
\end{equation}
which gives the dimensionless retraction time in the capillary-inertial regime $t_{r,c-i}^{\ast}$. On the other hand, in the capillary-viscous regime, the difference of surface energy of the drop at $t/t_r = 0$ and $t/t_r = 1$ is dissipated by predominant uniaxial-deformation viscous effects $[3 \eta {\dot{\gamma}}_r  (R_0-r)  \pi (R+r)^2]$, where the prefactor 3 emerges from the uniaxial extensional nature of the flow and ${\dot{\gamma}}_r = (k/t_r) ~ \ln(R_0/r)$ is the characteristic deformation rate associated with the capillary-viscous regime, as underlined by sub-figures \ref{fig-1}\textit{d}-\textit{f}]. Equating the surface and the dissipated energy leads to
\begin{equation}
t_r \left(  \frac{U_r}{2R_0}  \right) \approx  \mathrm{Ca} \frac{ 3k ~ ln(R_0/r) (R_0-r)  (R+r)^2  }{ 4 R_0^3 \left(  \frac{R^2}{R_0^2} + \frac{\pi R r}{R_0^2} - 2  \right)  }  =  t_{r,c-v}^{\ast}  \, ,
\label{eq:cap-visc-scaling}
\end{equation}
which denotes the dimensionless retraction time in the capillary-viscous regime $t_{r,c-v}^{\ast}$. Lastly, the transition between the regimes mentioned above can be found by equating our scaling laws (equations \ref{eq:cap-in-scaling} and \ref{eq:cap-visc-scaling}), which yields   
\begin{equation}
\frac{t_{r,c-v}^{\ast}}{t_{r,c-i}^{\ast}} = \frac{\mathrm{Ca}}{\mathrm{We}} \frac{ 9k ~ ln(R_0/r) ~ (R_0-r) ~ (R+r)^2  }{ R_0^2 ( R + r - R_0 )  } \approx 1  \, .
\label{eq:retractio-number}
\end{equation}
This equation indicates that the transition between the capillary-inertial and the capillary-viscous regime occurs at $t_{r,c-v}^{\ast} / t_{r,c-i}^{\ast} \approx 1$, where $t_{r,c-v}^{\ast} / t_{r,c-i}^{\ast}$ represents a dimensionless parameter called here the \textit{retraction number} \citep[analogous to the impact number reported by][]{Isukwem-24a, Isukwem-24b}. Hence, the retraction number gravitates around one in the mixed capillary-inertio-viscous regime (relative to the results shown in sub-figures \ref{fig-5}\textit{b}-\textit{c}, for instance). 

The validity of the above theoretical arguments is corroborated by figure \ref{fig-6}, in which the rescaled retraction time $t_r(U_r/2R_0)/t_{r,c-i}^{\ast}$ is plotted as a function of the retraction number $t_{r,c-v}^{\ast} / t_{r,c-i}^{\ast}$ (with $k=200$, which shift the mixed regime towards $t_{r,c-v}^{\ast} / t_{r,c-i}^{\ast} \approx 1$) for a variety of $D_{max}/D_0$ values (different symbols). The results collapse across a single path divided into three regions highlighting different retraction regimes: capillary-inertial (I; in blue; $t_{r,c-v}^{\ast} / t_{r,c-i}^{\ast} \leq 0.1$); capillary-viscous (II; in pink; $t_{r,c-v}^{\ast} / t_{r,c-i}^{\ast} \geq 10$); and mixed capillary-inertio-viscous (III; in white; $0.1 < t_{r,c-v}^{\ast} / t_{r,c-i}^{\ast} < 10$). Hence, flow cases associated with the blue region emerge from a competition between capillary and inertial stresses (similar to sub-figure \ref{fig-5}\textit{a}), whereas the flow cases related to the pink region rise from a balance between capillary and viscous stresses (see sub-figures \ref{fig-5}\textit{d}-\textit{h}). The points landing within the white region represent flow scenarios for which capillary, inertial and viscous effects are equally relevant (sub-figures \ref{fig-5}\textit{b}-\textit{c}, for instance).

Finally, it is important to note that, as anticipated by the energy-budget figure, $\Pi_{\mathrm{We}} \geq$ 0.9 and $\Pi_{\mathrm{Ca}} \leq$ 0.1 within the region I (blue; capillary-inertio regime) in figure \ref{fig-6}, while $\Pi_{\mathrm{We}} \leq$ 0.1 and $\Pi_{\mathrm{Ca}} \geq$ 0.9 in the region II (red; capillary-viscous regime). In the region III (white; mixed capillary-inertio-viscous regime), $ 0.1 < \Pi_{\mathrm{We}} < 0.9$ and $0.1 < \Pi_{\mathrm{Ca}} < 0.9$.    

%%%%%%%%%%%%%%%%%%%%%%%%%%%%%%%%%%%%%%%%%%%%%%%%%%%%%%%%%%%%%%%%%%%%%%%%%%%%%%
%%%%%%%%%%%%%%%%%%%%%%%%%%%%%%%%%%%%%%%%%%%%%%%%%%%%%%%%%%%%%%%%%%%%%%%%%%%%%%
\section{Concluding Remarks}
%%%%%%%%%%%%%%%%%%%%%%%%%%%%%%%%%%%%%%%%%%%%%%%%%%%%%%%%%%%%%%%%%%%%%%%%%%%%%%
%%%%%%%%%%%%%%%%%%%%%%%%%%%%%%%%%%%%%%%%%%%%%%%%%%%%%%%%%%%%%%%%%%%%%%%%%%%%%%
%\vspace*{-0.5cm}
We have presented here a theoretical and numerical study devoted to the physical mechanisms driving the retraction of Newtonian levitating drops following their spreading. Our 3D numerical simulations were based on a variational multi-scale approach dedicated to multiphase flows. The results obtained were analysed considering the retraction dynamics, energy balances and scaling laws. 

The retraction regimes have been highlighted: the capillary-inertial regime, resulting from a competition between capillary and inertial stresses, for which $t_r(U_r/2R_0) = \mathrm{We} \frac{R+r-R_0}{ 12 R_0 \left(\frac{R^2}{R_0^2} + \frac{\pi R r}{R_0^2} - 2 \right)}$; the capillary-viscous regime, resulting from a balance between capillary and viscous stresses, for which $t_r(U_r/2R_0) =  \mathrm{Ca} \frac{ 3k ~ ln(R_0/r) (R_0-r)  (R+r)^2  }{ 4 R_0^3 \left(  \frac{R^2}{R_0^2} + \frac{\pi R r}{R_0^2} - 2  \right) }$; and the mixed capillary-inertia-viscous regime, for which capillary, inertial and viscous stresses are all important. These results were synthesized in the form of a master curve giving the retraction time for levitating Newtonian drops as a function of a unique dimensionless number called retraction number $ \frac{\mathrm{Ca}}{\mathrm{We}} \frac{ 9k ~ ln(R_0/r) ~ (R_0-r) ~ (R+r)^2  }{ R_0^2 ( R + r - R_0 ) }$ relating the three important dimensionless parameters of the problem, namely $\mathrm{We}$, $\mathrm{Ca}$ and geometrical aspects of the drop $D_{max}/D_0 =(R+r)/R_0$. 

Interestingly, the stress balance driving the capillary-inertia regime induces a complex flow with equivalent shear, uniaxial, and biaxial components. As the viscous stress becomes comparable to the capillary and inertial ones, a mixed regime emerges while shear-based deformations tend to vanish. Lastly, flow cases under capillary-viscous dominance essentially exhibit axial deformation. 

Our study has direct applications in several flow situations promoting non-contact between Newtonian stretched drops and solid or liquid surfaces, such as Leidenfrost drops, bouncing drops, walking drops and levitating drops in acoustic fields or moving surfaces \citep{Leidenfrost-56, Hall-69, Garmett-78, Maquet-16, Adda-Bedia-16, Gauthier-19, Graeber-21, Wang-22, Isukwem-24d, Jayaratne-64, Gopinath-01, Richard-02, Bach-04, Zou-11, Molacek-13a, Tang-19, Couder-05, Molacek-13b, Bush-20, Hasegawa-19, Chen-24, Lhuissier-13}. Since complex fluids are frequently used in industrial configurations, it would be interesting to consider non-Newtonian effects on the retraction of levitating drops, such as those related to yield stress, thixotropy and/or elasticity. 

%%%%%%%%%%%%%%%%%%%%%%%%%%%%%%%%%%%%%%%%%%%%%%%%%%%%%%%%%%%%%%%%%%%%%%%%%%%%%%
%%%%%%%%%%%%%%%%%%%%%%%%%%%%%%%%%%%%%%%%%%%%%%%%%%%%%%%%%%%%%%%%%%%%%%%%%%%%%%
\textit{\textbf{Acknowledgements:}} 
%%%%%%%%%%%%%%%%%%%%%%%%%%%%%%%%%%%%%%%%%%%%%%%%%%%%%%%%%%%%%%%%%%%%%%%%%%%%%%
%%%%%%%%%%%%%%%%%%%%%%%%%%%%%%%%%%%%%%%%%%%%%%%%%%%%%%%%%%%%%%%%%%%%%%%%%%%%%%
The authors would like to acknowledge the support from the UCA$^\mathrm{JEDI}$ program (IDEX of Universit\'e C\^ote d'Azur), the PSL Research University under the program `Investissements d'Avenir' launched by the French Government and implemented by the French National Research Agency (ANR) with the reference ANR-10-IDEX-0001-02 PSL, and the ANR for supporting the INNpact project under the `Jeunes chercheuses et jeunes chercheurs' program.   

%\newpage
%\clearpage

%%%%%%%%%%%%%%%%%%%%%%%%%%%%%%%%%%%%%%%%%%%%%%%%%%%%%%%%%%%%%%%%%%%%%%%%%%%%%%
%%%%%%%%%%%%%%%%%%%%%%%%%%%%%%%%%%%%%%%%%%%%%%%%%%%%%%%%%%%%%%%%%%%%%%%%%%%%%%
\bibliographystyle{./apsrev4-2}
\bibliography{Retraction-ref}% Produces the bibliography via BibTeX.
%%%%%%%%%%%%%%%%%%%%%%%%%%%%%%%%%%%%%%%%%%%%%%%%%%%%%%%%%%%%%%%%%%%%%%%%%%%%%%
%%%%%%%%%%%%%%%%%%%%%%%%%%%%%%%%%%%%%%%%%%%%%%%%%%%%%%%%%%%%%%%%%%%%%%%%%%%%%%
\end{document}